\documentclass{aa}
\usepackage{graphicx,epsf,epsfig}

\usepackage{txfonts}
\begin{document}
   \title{1RXS\,J173021.5-055933: a cataclysmic variable with a 
fast-spinning magnetic white dwarf\thanks{Based on observations obtained 
with XMM-Newton and INTEGRAL, ESA
science missions with instruments and contributions directly funded by ESA 
Member States and NASA}}


   \author{D. de Martino
          \inst{1}
          \and
          G. Matt\inst{2}\and K. Mukai\inst{3}\and
J.-M. Bonnet-Bidaud\inst{4}, M. Falanga\inst{4}, B.T. G\"ansicke\inst{5},
F. Haberl\inst{6} \and T.R. Marsh\inst{5}, M. Mouchet\inst{7}, S.P. 
Littlefair\inst{8}, V. Dhillon\inst{8}
 }

\offprints{D. de Martino}

   \institute{INAF--Osservatorio Astronomico di Capodimonte, Via
Moiariello 16, I-80131 Napoli, Italy\\
              \email{demartino@na.astro.it}
         \and
Dipartimento di Fisica, Universita' degli Studi Roma Tre, Via della
Vasca Navale 84, I-00146 Roma, Italy \\
\email{matt@fis.uniroma3.it}
\and
CRESST and X-ray Astrophysics Laboratory NASA/GSFC, Greenbelt,
MD 20771 and Department of Physics, University of Maryland, Baltimore 
County, 1000 Hilltop Circle, Baltimore, MD 21250, USA\\
 \email{mukai@milkyway.gsfc.nasa.gov}
\and
Service d'Astrophysique, DSM/DAPNIA/SAp, CE Saclay, F-91191 Gif sur Yvette
Cedex, France\\
\email{bonnetbidaud@cea.fr,mfalanga@cea.fr}
\and
Department of Physics, University of Warwick, Coventry CV4 7AL,
UK\\ 
\email{boris.gaensicke@warwick.ac.uk,Tom.Marsh@warwick.ac.uk}
\and 
Max-Planck-Institut f\"ur Extraterrestrische Physik,
Giessenbachstra{\ss}e, Postfach 1312, 85741 Garching, Germany \\
\email{fwh@mpe.mpg.de}
\and
Laboratory APC, University Denis Diderot, 10 Rue Alice Domon et Leonie 
Duquet, F-75013 and 
LUTH, Observatoire de Paris, F-92190 Meudon, France\\
\email{martine.mouchet@obspm.fr}
\and
Department of Physics and Astronomy, University of Sheffield, Sheffield S3 
7RH, UK\\
\email{s.littlefair@shef.ac.uk,vik.dhillon@shef.ac.uk}
             }

   \date{Received July 30, 2007; accepted January 20, 2008}

\abstract{}
{We present the first X-ray observations with the XMM-Newton and 
INTEGRAL satellites of the recently discovered 
cataclysmic variable 1RXS\,J173021.5-055933, together with simultaneous 
UV and coordinated optical photometry aiming at characterising its 
broad-band  temporal and spectral properties and classifying 
this system as a magnetic one.}
{We performed a timing analysis of the X-ray, UV, and optical light curves
to identify and to study the energy dependence of the fast 
128\,s  pulsation over a wide energy range. X-ray spectral analysis 
in the  broad 0.2-100\,keV X-ray range was performed to characterise 
the peculiar emission properties of this source.  
}
{We find that the X-ray light curve is dominated by the spin period of 
the accreting white dwarf in contrast to the far-UV range, which turns out 
to be  unmodulated at a 3$\sigma$ level. Near-UV and optical 
pulses are instead detected at 
twice the 
spin frequency. We identify the contributions from  two accreting 
poles that imply a moderately inclined dipole field allowing, one pole 
to dominate at energies at least up to 10\,keV, and  a secondary 
that instead is negligible above 5\,keV. 
X-ray spectral analysis 
reveals the presence of multiple emission components consisting of  
optically thin plasma  with temperatures 
ranging from 0.17\,keV
to 60\,keV and a hot blackbody  at $\sim$90\,eV. 
The spectrum is also strongly affected by peculiar absorption 
components consisting of two 
high-density ($\rm \sim 3\times 10^{21}\,cm^{-2}$ and $\rm 2\times 
10^{23}\,cm^{-2}$)  intervening columns, plus a warm absorber. The last 
is detected from an  OVII absorption  edge at 0.74\,keV, which suggests 
that photoionization of pre-shock material is also occurring in this 
system.
} 
{The observed properties indicate that the 
accretor in 1RXS\,J173021.5-055933 is a white dwarf with a likely weak
magnetic field, thus confirming this cataclysmic variable as an 
intermediate polar (IP) with one of the most extreme spin-to-orbit period 
ratios. This system also joins the small group of IPs 
showing a soft X-ray reprocessed component, suggesting that this 
characteristics is not uncommon in these systems.  
}


   \maketitle
%

\section{Introduction}

Cataclysmic variables (CVs) are close binaries containing an accreting
white dwarf (WD) from a late type Roche-lobe filling (main sequence or
subgiant) secondary star (for a comprehensive review see Warner 
\cite{Warner95}).  
Among them, the magnetic systems (mCVs) nowdays constitute
a rather conspicuous group ($\sim$25$\%$ of all 
CVs).  The current census of mCVs, including confirmed ones and 
candidates, 
amounts to about 150 systems (Ritter \& Kolb (\cite{RitterKolb06}). They are further
divided in two classes:  the polars, containing a strongly magnetised
($>$10\,MG) WD, whose rotation is synchronised with the
orbital period and the intermediate polars (IPs),  which contain 
accreting WDs that are rotating much 
faster than the orbital period. The presence of X-ray pulses at the WD 
rotation but the lack of detectable 
polarized circular radiation in the optical/near-infrared in most systems, 
led to believe that IPs possess weakly magnetised WDs (B$\lesssim $10\,MG) 
(see review by Patterson \cite{patterson}). However this is 
still the subject of great debate (see below).
As of today, the polars consist of $\sim$58$\%$ of the 
known mCVs. Because of their strong soft
X-ray ($\sim$40\,eV) emission component, the polars increased in number
after the ROSAT era (Beuermann \cite{Beuermann99}).  This was not the 
case for the IPs, which are  dominated  by a hard 10-20\,keV optically 
thin 
emission.
However in the last few years a large number ($\sim$34) of IP candidates 
were  discovered from ground-based observations (Warner \& Woudt 
\cite{WarnerWoudt04}, Rodriguez-Gil et al.
\cite{Rodriguez-Gil05}, G\"ansicke et al. \cite{Gaensicke05}, 
Bonnet-Bidaud et al. \cite{BonnetBidaud06}, 
Southworth et al. \cite{Southworth06}, 
Southworth et al. \cite{Southworth07}).   Also, hard X-ray surveys 
such as those carried out by the {\em INTEGRAL} and {\em SWIFT} satellites are 
providing 
new and interesting results on CVs 
(Wheatley et al. \cite{Wheatley2006}, Masetti et al. 
\cite{masetti06a}, Masetti et al. \cite{masetti06b}). In particular, 
from the 
current {\em INTEGRAL} catalogue (Bird et al. \cite{Bird07}) about 5$\%$ 
of the detected sources are CVs, most of them are 
IPs (Barlow et al. \cite {Barlow06},  Falanga et al. \cite{falanga}). This 
has  further stimulated  optical and X-ray follow-ups to 
search for new IPs (e.g. Bonnet-Bidaud et al. 
\cite{BonnetBidaud07}). The detection of high energy emission of IPs 
extending up to  $\sim$90\,keV was already identified from {\em BeppoSAX} 
observations of a few bright systems (de Martino et al. 
\cite{demartino01}, de Martino et al. \cite{demartino04}), which makes the 
new discoveries in the hard X-rays very promising to understand the 
role of this, formerly, small group of mCVs.

The link between polars  and IPs, and hence the evolution of mCVs, is 
still uncertain.
Polars are typically found at short orbital periods, most of them below
the so-called 2-3\,hr orbital period gap. IPs instead populate
the orbital period distribution at longer periods with a handful of
systems below the gap. The majority of IPs have rotational periods 
($\rm P_{spin}$) about 7-10$\%$ the binary orbital periods ($\rm 
P_{orb}$). However, including the new optical candidates  a  wide 
range of degree 
of asynchronism,  $\rm P_{spin}/P_{orb}$ from $\sim$0.002 to 
$\sim 0.7$, is encountered (Woudt \& Warner \cite{WarnerWoudt04}, de 
Martino et al. \cite{demartino06a}).  This might suggest that 
the former  separation between IPs and polars could be only apparent and
reinforces the debate on the hypothesis that IPs possess  
similar magnetic fields to those of the low-field polars, and will 
synchronise 
while evolving towards short orbital periods (Cumming \cite{Cumming}, 
Norton et al. \cite{norton04}, Southworth et al. \cite{Southworth07}). 
The confirmation of new optically discovered candidates 
is therefore essential in this respect. Differently
from the optical range, heavily affected by X-ray reprocessing, a
secure identification comes from the detection of X-ray
pulses at the WD spin period, which arise from the post-shock region of
the magnetically confined accretion flow onto the WD.  

In the framework of a programme with {\em XMM-Newton}
aiming at unambiguously confirming the IP nature of new optically
identified systems, we here present the first X-ray study of 
1RXS\,J173021.5-055933 (hereafter RXJ\,1730). Identified as a magnetic CV
by  G\"ansicke et al. (\cite{Gaensicke05}), its  
very fast (128\,s)  optical pulsations and very long 
(15.4\,hr)  orbital period make it an extreme object in terms  
of degree of asynchronism.  RX\,J1730 has the second fastest spinning WD 
preceeded only by AE\,Aqr with a spin period of 33\,s. Its orbital 
period is the second longest ever observed in mCVs, being headed by  
GK\,Per  with a period of 47.9\,hr.  While GK\,Per and AE\,Aqr likely 
harbour  K subgiant secondary stars 
(Morales-Rueda et al. \cite{morales-rueda}, Casares et al. \cite{casares}) 
the  donor star in RX\,J1730 was suggested  to be a G-type  main sequence 
star, and differently 
from these two, the optical spectrum is dominated by the accretion flow 
rather than by  the donor star (G\"ansicke et al. \cite{Gaensicke05}).  
The  nature and evolutionary path of very long  
orbital period systems is still unknown though of great importance to 
understand the evolution of mCVs as a whole.

With the purpose of identifying unambiguously the rotational period of the 
WD primary   and  to study the emission and accretion properties of 
this system, the analysis 
presented here include a temporal and spectral analysis of the   
{\em XMM-Newton} observation. It was also detected in the hard X-rays
during the IBIS/ISGRI galactic plane survey as
IGR\,J17303-0601=INTEGRAL1 56 (Bird et al. \cite{Bird04}) and formerly 
proposed to  be a low mass X-ray binary (LMXB) by Masetti et al. 
(\cite{masetti04}). Hence, we included in 
the  spectral  analysis also the  archival data from the {\em 
INTEGRAL} satellite. We further   complement the X-ray study with 
simultaneous far-UV photometry  from the  {\em XMM-Newton} Optical Monitor 
as well as coordinated quasi-simultaneous near-UV/optical photometry 
acquired at the {\em William Herschel Telescope} in La Palma.

   \begin{table*}[t!]
      \caption{Summary of the observations of RX\,J1730.}
         \label{obslog}
     \centering
\begin{tabular}{l  c c r c c}
            \hline \hline
            \noalign{\smallskip}
Instrument     &  Date & UT(start) & Duration (s) & Net Count Rate (cts\,s$^{-1}$)\\
            \noalign{\smallskip}
            \hline
            \noalign{\smallskip}
{\em XMM-Newton} & & & & \\
 EPIC PN &  2005 Aug 29 & 07:05 & 11271 & 3.73\\
 EPIC MOS & & 06:37 & 13275 & 1.02 \\
 RGS    & & 06:36 & 13507 & 0.12  \\
 OM UVW1 & & 06:45 & 2319 & 2.57\\
      & & 07:29 & 2320 & \\
      & & 08:13 & 2320 & \\
      & & 08:58 & 2320 & \\
      & & 09:42 & 2319 & \\
            \hline
                 \noalign{\smallskip}
Instrument     &  Date & N. Science Window & Duration$^a$ (s) & Net Count 
Rate (cts\,s$^{-1}$)\\
            \hline
                 \noalign{\smallskip}
{\em INTEGRAL}/IBIS & & & & 0.57\\
       & 2003 Apr 23-30 & 75 & 135000 & \\
           & 2003 May 1 & 4 & 9182 & \\
           & 2003 Aug 30 & 5 & 17757 & \\
           & 2003 Sept 3 & 3 & 8637 &\\
           & 2003 Oct 10-17 & 95  & 172000 & \\
           & 2004 Feb 15-25 & 28 & 61800 & \\
           & 2004 Mar 11-20 & 35 & 66700 & \\
           & 2004 Apr 2     &  6 & 19875 & \\
 \hline
                 \noalign{\smallskip}
Instrument     &  Date & UT(start) & Duration (s) & r'\,(mag)\\
            \noalign{\smallskip}
            \hline
            \noalign{\smallskip}
{\em William Herschel Telescope} & & & & \\
ULTRACAM/u',g',r' & 2005 Aug 27 & 20:13 & 2567 & 15.6 \\
         & 2005 Aug 28 & 20:30 & 1925 & 15.6 \\
         & 2005 Aug 29 & 20:15 & 3476 & 15.4 \\
         & 2005 Aug 30 & 20:19 & 1255 & 15.3 \\
         & 2005 Aug 31 & 20:26 & 992  & 15.3 \\
         & 2005 Sep 1 &  20:11 & 1640 & 15.3 \\
            \noalign{\smallskip}
            \hline \hline
\end{tabular}
\begin{flushleft}
$^a$: Total exposure time
\end{flushleft}
\end{table*}


\section{The observations}
 
\subsection{The XMM-Newton observation}

RX\,J1730 was observed with the {\em XMM-Newton\/} satellite (Jansen et
al. \cite{jansenetal}) on 2005 Aug. 29
(obsid:0302100201). The EPIC-PN (Str\"uder et
al. \cite{strudeetal}) and MOS (Turner et al. \cite{turneretal}) cameras
were operated in large window mode   with 
the medium filter for net
exposure times of about 11.3\,ks and 13.3\,ks respectively. RX\,J1730 was 
also observed with the Reflection Grating Spectrographs (RGS1 and RGS2) 
(den Herder et al. \cite{denherderetal}) in
spectroscopy mode with an exposure time of
13.5\,ks and with the Optical Monitor (OM) instrument (Mason et
al. \cite{masonetal}) using the wide UVW1 filter
covering the range 2450--3200\,\AA \, (effective wavelength 2910\,\AA 
\,), in imaging fast mode for a total exposure time of
11.6\,ks. A summary of the observations is reported in Table~\ref{obslog}.

All data sets were reprocessed using the {\em XMM-Newton} SAS 6.5 package 
using the most updated calibration files.  
The EPIC light curves and spectra were extracted 
within the same SAS package from a circular region 
with a radius of 25$^{"}$ centred on the source. Background light curves 
and spectra were extracted from  offset circular regions with same radii 
as for the target on the same CCD chip.  
Single and double pixel events with a
zero quality flag were selected for the EPIC-PN data, while for EPIC-MOS
cameras up to quadruple pixel events were  used. 
The RGS pipeline was run using the SAS task {\em rgsproc}.

A long background flaring activity  occurred 6.2\,ks after the start of 
the {\em XMM-Newton\/}  observation. The average count rate during 
the flare 
is 0.13\,cts\,s$^{-1}$
and 0.02\,cts\,s$^{-1}$ in the EPIC-PN and MOS cameras respectively, reaching
0.6\,cts\,s$^{-1}$ (PN) and 0.2\,cts\,s$^{-1}$ (MOS) at flare maximum. 
Because the source was on average 26 and 10 times brighter than 
 the  background 
during the flare event does in both cameras respectively, 
the light curves are not greatly affected. 
On the other hand, since the source of flares are solar protons 
with unpredictable spectral 
behaviour,  we conservatively limit the spectral analysis of the EPIC data 
to the first  6.2\,ks of observation. Also in 
the RGS instruments the count rate increases during flare up to $\rm \sim 
1\,cts\,s^{-1}$. However the soft  protons are not dispersed by the 
gratings and a  comparison of 
the RGS spectra extracted including and without the flare periods, 
shows that the spectra are indeed not affected by this event. We 
therefore use for the RGS spectral analysis the whole observation.
The  EPIC-PN and MOS and the RGS1 and RGS2 
first order spectra were rebinned to have 
 a minimum of 20 counts in each bin. 
Ancillary response and redistribution matrix files were created using SAS
tasks {\em arfgen} and {\em rmfgen} respectively.

OM background subtracted light curves were also produced using SAS from 
the reprocessed event files, extracting the source count rates from a 
circular region of $\sim 3^{"}$ radius and the backgouund rates from 
an  outer annulus with $\delta r \sim 2^{"}$ centred on the source. 
Average net count rate in the UVW1 filter is 
2.57\,cts\,s$^{-1}$, 
corresponding to an instrumental magnitude of 
16.3\,mag. Using Vega magnitude to flux conversion, this
corresponds to a flux of $\rm 1.2\times 
10^{-15}\,erg\,cm^{-2}\,s^{-1}\,\AA^{-1}$
in the 2450-3200\,\AA \, band.  When compared  with the
extrapolation of the optical (4000--7000\,\AA \,) spectra obtained by 
G\"ansicke et al. (\cite{Gaensicke05}) at different epochs, the UV flux is 
higher than that observed in June 2003 but lower than that of April 2003. Heliocentric corrections were 
applied to the extracted 
light curves. 

\subsection{The INTEGRAL data}

RX\,J1730 was detected as a hard X-ray source in the  {\em INTEGRAL}  
(Winkler et al. \cite{w03}) 
IBIS/ISGRI  (Ubertini et al. \cite{u03}, Lebrun et al. \cite{lebr03}) 
soft gamma-ray Galactic Plane 
Survey (Bird et al. \cite{Bird04}) with a count rate of 
0.28$\pm$0.03\,cts\,s$^{-1}$ in  the 20--40\,keV range. 
As it will be shown in Sect.\,3, the X-ray  emission of RX\,J1730 is 
hard, with substantial emission above 10\,keV. Hence, to extend the study 
to 
higher energies we analysed all public IBIS/ISGRI coded 
mask data in 
the range 20--100\,keV. Several observing windows were  
included spanning from April 2003 to 
April 2004 (see Table~\ref{obslog}).  The data were extracted for all 
ISGRI pointings with a source
position offset $\leq 9^{\circ}$ for a total effective exposure of $\sim 340$\,ks.
Because of the weakness of the source and several short time windows, the 
data were only used  to study the X-ray spectrum in the 20--100\,keV 
range. 
The ISGRI spectrum was extracted using  the standard  {\em INTEGRAL} Offline Science Analysis (OSA)  
software version 5.1.
Single pointings were deconvolved and analysed separately, and then 
combined in mosaic images.
 The source is clearly detected  with a net count rate of 0.57\,cts\,s$^{-1}$
 in the energy band 20--80 keV, 
at a significance level of $13.6\sigma$. At energies above 100 keV
RX\,J1730 was not detected at a statistically significant level in the total
exposure time.

\subsection{The optical photometry}

RX\,J1730 was observed for six consecutive days at the 4.2\,m 
{\em William Herschel Telescope} (WHT) at La Palma equipped with 
ULTRACAM (Dhillon et al. \cite{Dhillon}), which provides 
imaging photometry at high temporal resolution in three different colours 
simultaneously. The observations 
were performed from Aug. 27 to Sept. 1, 2005 using the Sloan Digital 
Sky Survey (SDSS) u',g',r' 
filters with a temporal resolution of about 2\,s. The log of the observations 
is reported in Table~\ref{obslog}, together with the average target 
magnitude level for each night. The seeing was variable in quality as 
well as transparency. 
The data were reduced using the ULTRACAM pipeline data-reduction 
system. Aperture pho\-to\-metry was performed on RX\,J1730 and several 
comparison stars.   
Differential photometry in the three bands was obtained by ratioing the 
background subtracted count rates of the target and best comparison star. 
Heliocentric correction was applied to the light curves in the three bands.

   \begin{figure}
   \centering
\includegraphics[height=9.cm, width=8.cm,angle=0]{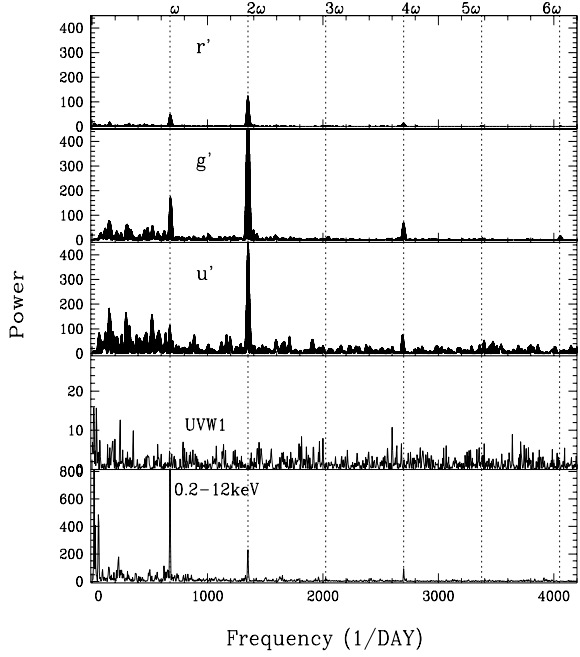}
\caption{Power spectra of RX\,J1730.  
From bottom to top: the 
EPIC-PN power spectra in 0.2-12\,keV 
range, in the UV and u',g',r' bands,  using light curves of 
Aug. 27, 28, and 29 (see text for details).}\label{fig1}
    \end{figure}

\section{Timing analysis}

\subsection{Search for periodicities}

\begin{figure}[h!]
   \centering
\includegraphics[height=9.cm, width=8.cm,angle=0]{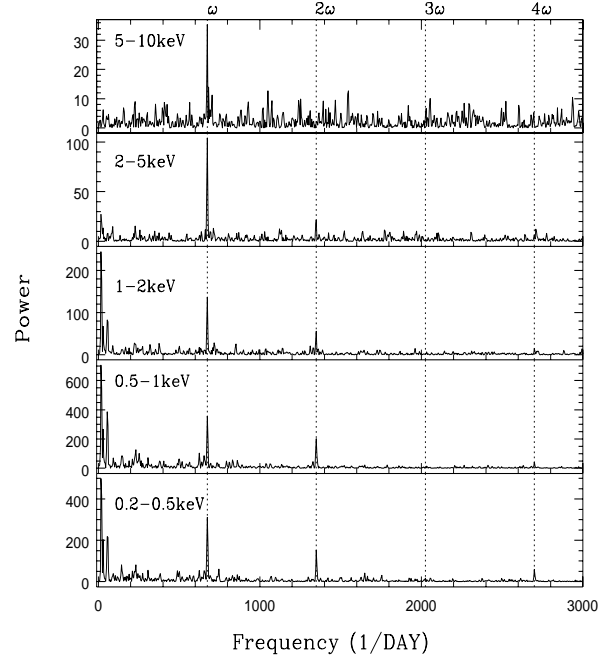}
\caption{EPIC-PN power spectra in selected energy ranges. {\em From bottom 
to top:} 0.2--0.5\,keV, 0.5--1\,keV, 1--2\,keV, 2--5\,keV and
5--10\,keV. The spin ($\omega$) and its harmonics  are marked with vertical
dotted lines. The fundamental dominates at all energies while the first harmonic 
is weak in the harder bands.}
\label{fig2}
    \end{figure}

To search for periodic variations in the X-ray band we extracted light
curves in the 0.2-12\,keV range from the EPIC PN and MOS cameras with a
time resolution of 2\,s. A Fourier analysis shows a strong peak at
675\,day$^{-1}$ as well as at its first and third harmonic confirming the
optical periodicity found by G\"ansicke et al. (\cite{Gaensicke05}). 
This is shown in the bottom panel of Fig.~\ref{fig1} for the EPIC-PN
light curve. The EPIC-MOS light curves are noisier but confirm this
result.  Hence the X-ray  128\,s periodicity is confirmed to be the 
rotational  period of the accreting WD.   
 A sinusoidal fit to the EPIC-PN light curve with three sinusoids
accounting for the fundamental, the first and third harmonics gives 
$\rm P_{\omega}=128.02\pm 0.02$\,s and  the
following ephemeris for the time of minimum of X-ray pulse: $\rm
HJD_{min}^{X}$=2\,453611.86477(1))+0.0014817(2)\,E. 
We also performed a
Fourier analysis on the extracted (10\,s) light curves in selected energy
bands as shown in Fig.~\ref{fig2}. The fundamental dominates at all
energies, while the first harmonic gets weaker at higher energies,
indicating that the X-ray pulse is more sinusoidal above 2\,keV.

   \begin{figure}[h!]
   \centering
\includegraphics[height=9.cm, width=8.cm,angle=0]{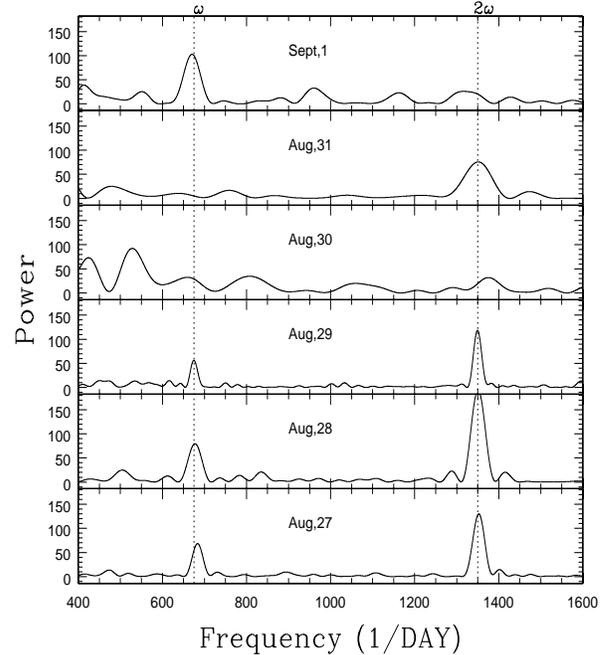}
\caption{Power spectra of the r' band light curves of RX\,J1730 from Aug. 
27 to 
Sept. 1, showing strong changes in the amplitude of the spin 
variability. Power at the fundamental frequency changes more than 
a factor of 2 from Aug. 31 to Sept. 1. Ordinates are arbitrary.}
\label{fig3}
    \end{figure}

A search for periodic signals in the UVW1 light 
curve extracted with a  temporal resolution  of 2\,s surprisingly did 
not  show   any variability above the 3$\sigma$ level (Fig.~\ref{fig1}). 
This was also checked in each of the five UV data sets. 
We also inquired about any instrumental problem in 
the OM observation but no anomalies were found. Hence the lack of detectable 
variability in the UVW1 filter should be regarded as real  (but see also 
below). 
 
The u', g', and r' light curves were also Fourier analysed 
to search for short periodic signals. The long term trend in each night 
was removed using a third order polynomial. In Fig.~\ref{fig3}
we show the  r' band power spectra for each night revealing that the 
128\,s modulation is present, but absent on Aug. 30. Furthermore 
while from  Aug. 27 to Aug. 29 the first harmonic dominates over the 
fundamental and appears again on Aug. 31,  on  Sept. 1 only the 
fundamental is present. Strong and almost similar power 
at both fundamental and first harmonic was instead detected by G\"ansicke 
et al. (\cite{Gaensicke05}). Changes from night to night or 
cycle to cycle of the  amplitude of optical pulsations are not 
untypical of IPs (see Warner  \cite{Warner95} and references therein). The 
implications of the observed  behaviour in the optical will be discussed 
in Sect.\,5.  

The {\em XMM-Newton\/} observation was carried out 10.5\,hr after
the {\em  WHT\/}/ULTRACAM run of Aug. 28 and $\sim$13\,hr before that of 
Aug. 29. 
Although  further changes in the optical power spectrum cannot be 
excluded, the dominance of the first harmonic from Aug. 27 to Aug.29 
allows us to assume that the optical behaviour is constant in these three 
days. We then assume that the  X-ray power spectrum represents the 
behaviour of the X-ray source during that period, thus implying that the 
spin modulation is different in the two energy ranges.
Fig.~\ref{fig1} then reports a comparison between the X-ray and UV 
power spectra of Aug. 29 and those obtained from merging the optical 
u',g',r' 
band ones of Aug. 27, 28, and 29. Interesting to note is the presence of 
the  third harmonic not only in the X-rays but also in the optical and
best visible in the u' and g' bands. In addition and contrary to the 
X-rays,  power at the  fifth harmonic is detected in the g' band, but not 
at the fourth. 
The  fundamental  frequency 
is below the 3$\sigma$ level in the u' band, but re-emerges in the 
g' and r' bands. Also, the power at the first harmonic is similar in the u' 
and g' bands while it has lower amplitude modulation in the red. 

The detrended optical light curves of the first three nights were 
fitted with a  composite function consisting of three sinusoids at the 
fundamental, first and third  harmonics. For the g' band light curve, 
which 
provides the best fit, we obtain:  $\rm P_{\omega}=127.9999\pm 
0.0023$\,s  and an ephemeris for the time of minimum of the 
optical spin  pulse: 
$\rm HJD_{min}^{opt}$=2\,453611.746278(4)+0.00148148(3)\,E.

   \begin{figure*}[t!]
   \centering
\mbox{\epsfxsize=8cm\epsfysize=8cm\epsfbox{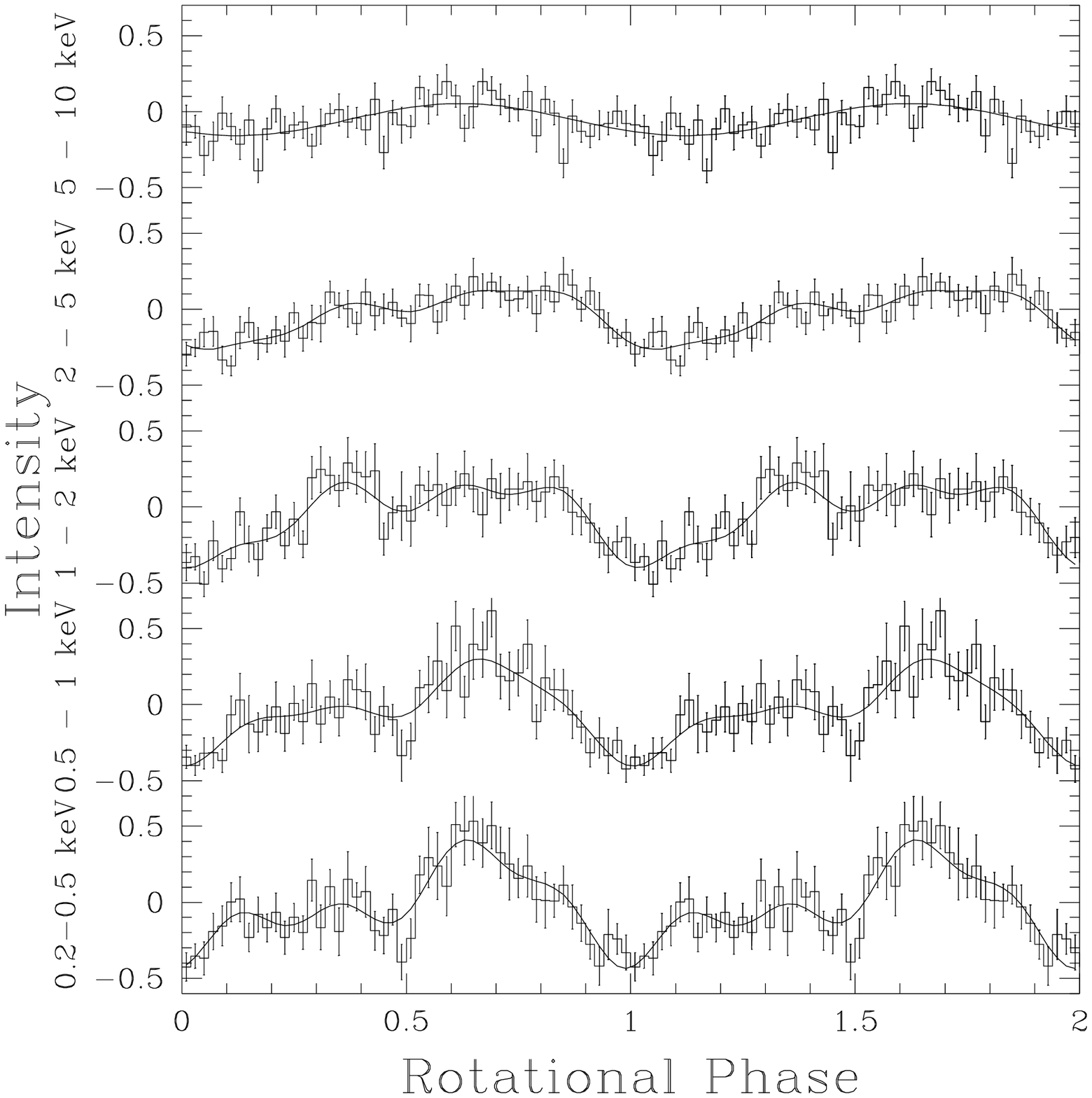}
\epsfxsize=8.cm\epsfysize=8cm\epsfbox{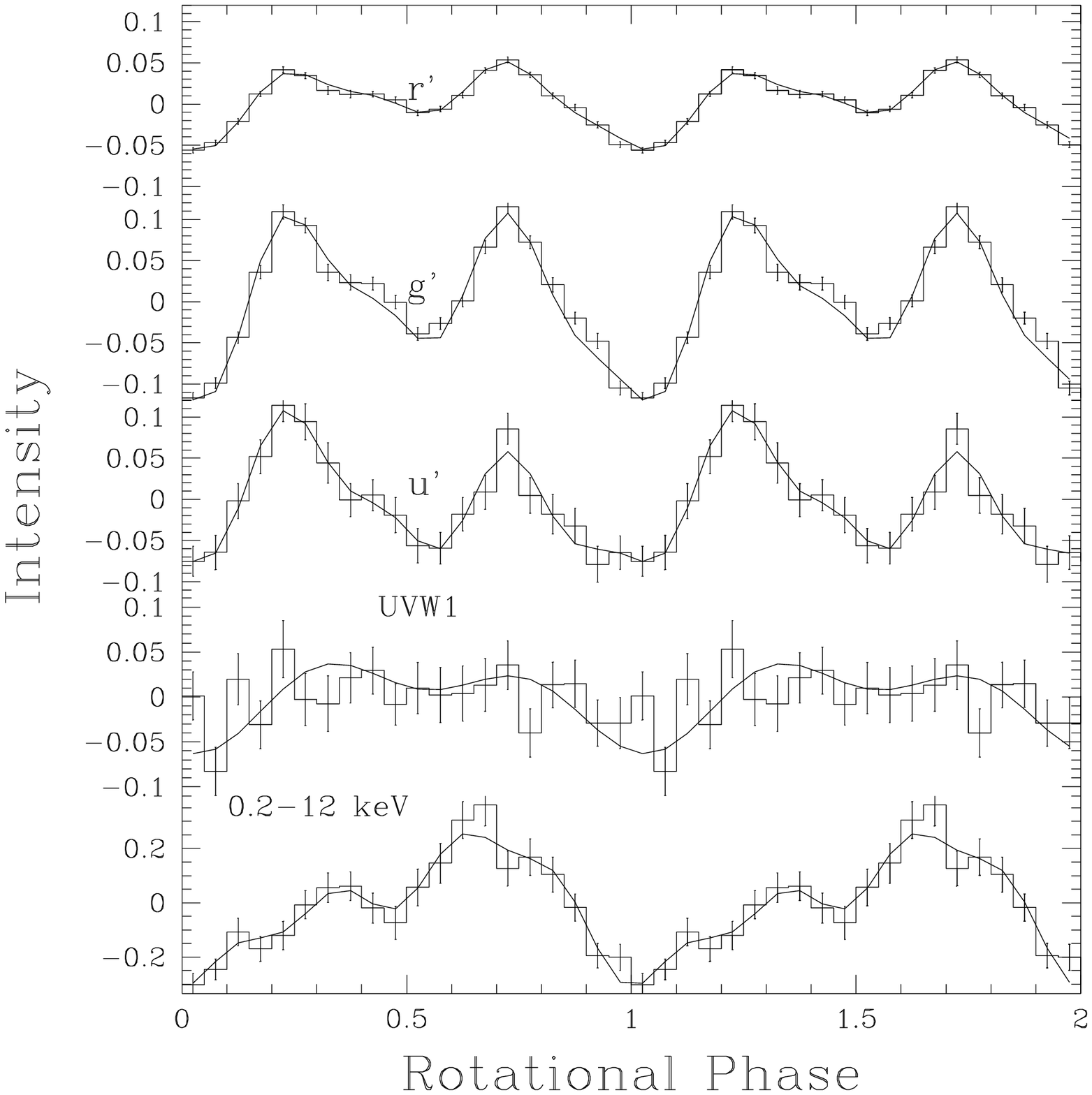}}
\caption{ {\em Left:} EPIC-PN folded light curves in selected energy 
bands at the 128\,s period using the optical ephemeris quoted in the text. 
{\em Right:} A comparison of the X-ray, UV 
and optical, u',g',r', band spin light curves. Ordinates are 
modulation intensities. Sinusoidal fits consisting of three sinusoids at 
the fundamental frequency, the first and third harmonics are also 
reported, except for the 
hardest band (5-10\,keV) where only one sinusoid at the fundamental period 
is used, and the UVW1 band where two sines were used (see 
text)}\label{fig4}
    \end{figure*}

\subsection{The spin pulsation}

To compare the X-ray and optical spin pulsations we use the more accurate
optical ephemeris for folding purposes.  In Fig.~\ref{fig4} (left panel),
we report the spin light curves in 50 phase bin points in different X-ray
energy bands together with the best fit composite function with three
sinusoids, at the fundamental, first and third harmonic frequencies
(except for the 5-10\,keV band, where only the fundamental is fitted).  
The amplitude of the modulation varies with energy being larger in
the soft X-rays. In particular the spin pulse full amplitudes are:  
48$\pm2\%$ in the 0.2-0.5\,keV band, 42$\pm2\%$ between 0.5-1\,keV,
24$\pm1\%$ in the 1-2\,keV and in the 2-5\,keV bands, and 10$\pm1\%$ in
the 5-10\,keV range. The decrease in modulation amplitude with increasing
energy is typically observed in IPs. Above 
2\,keV the pulse has a broad maximum at $\rm \phi_{rot}\sim$0.65 where 
it softens, while it gets harder at minimum. This behaviour is 
consistent with photoelectric absorption in the accretion flow. 
The pulsation gets more structured below 2\,keV as indicated by the 
presence of higher harmonics in the power spectra.

\noindent The ratios of the amplitudes  $\rm 
A_{\omega}\sim 1.4A_{2\omega}\sim 2.4A_{4\omega}$ in the 0.2-1\,keV 
range and   $\rm A_{\omega}\sim1.6A_{2\omega}\sim 4A_{4\omega}$ in the 
1-5\,keV  range,  indicate that there is  another emitting region 
offset by 180$^{\circ}$, contributing
$\sim66\%$ below 5\,keV. The amplitude of the third 
harmonic is larger in the softest band suggesting either additional 
absorption or a contribution from a much softer region.
Indeed and especially below 1\,keV, a narrow dip at $\rm \phi_{rot}\sim$ 
0.5 and a shallower one at $\rm \phi_{rot}\sim$ 0.2 can be recognised.  
 The last is best seen in the soft 0.2--0.5\,keV range. Since 
as a hardening is detected at these phases, this suggests additional absorption 
effects at $\rm \phi_{rot}\sim$ 0.2.

For the comparison of the X-ray and optical spin pulses, we 
again used the detrended u',g',r' band light curves of Aug.27, 28, and 
29. The folded light curves in the X-rays  and  optical bands are shown 
in  the right panel of Fig.~\ref{fig4}, together with the best sinusoidal 
fits 
that include three sinusoids with frequencies corresponding to the 
fundamental, the first and third harmonics.  For the UV light 
curve 
we also show a two-sine fit, though the amplitudes are below the 
3$\sigma$ level. Although the primary 
minimum occurs at about the same phases in both X-ray and optical bands, the 
pulse shape is different. The dip observed in the X-rays 
at $\rm \phi_{rot}\sim$0.5 appears as a pronounced secondary minimum in 
the optical.  The additional dip seen in the X-rays at 
 $\rm \phi_{rot}\sim$ 0.2 is not observed in the optical where instead we 
observe a second maximum with a shoulder at $\rm \phi_{rot}\sim$ 0.35.  
The optical pulsed fraction is about the same in u' 
($13.5\pm0.7\%$) and g'  bands ($16.2\pm0.5\%$), while it is $7\pm0.1\%$
in the r' band.  This last is consistent with that  observed by 
G\"ansicke et al. (\cite{Gaensicke05}).  On the other hand, the 
amplitude in the UVW1 band is $7.1\pm 2.6\%$ and hence a hardly detectable 
modulation in this range.
The differences of the optical spin pulse in the three bands are 
essentially  due to the different  amplitudes of first and third 
harmonics, being $\rm A_{\omega}\sim  0.5A_{2\omega}\sim1.8A_{4\omega}$ in 
the u' and g' bands, while in the r'  band  $\rm A_{\omega}\sim 
0.6A_{2\omega}\sim2.4A_{4\omega}$.  As a further comparison, taking at 
face values the amplitudes of the fundamental and first harmonic in the UVW1 
band, $\rm A_{\omega}\sim  1.4 A_{2\omega}$  are rather similar to what 
observed in the X-rays. This could suggest that the overall UV/optical 
pulse is not very blue and that only longwards 3000\,$\AA$, it is 
dominated by two equally contributing regions. We also 
note an asymmetry in the optical pulse  
profile at $\rm \phi_{rot}\sim$ 0.35  suggesting  
an additional  contribution of $\sim$26$\%$ 
 to  the optical  modulation. This will be discussed in 
Sect.\,5.

   \begin{figure*}[t,h!]
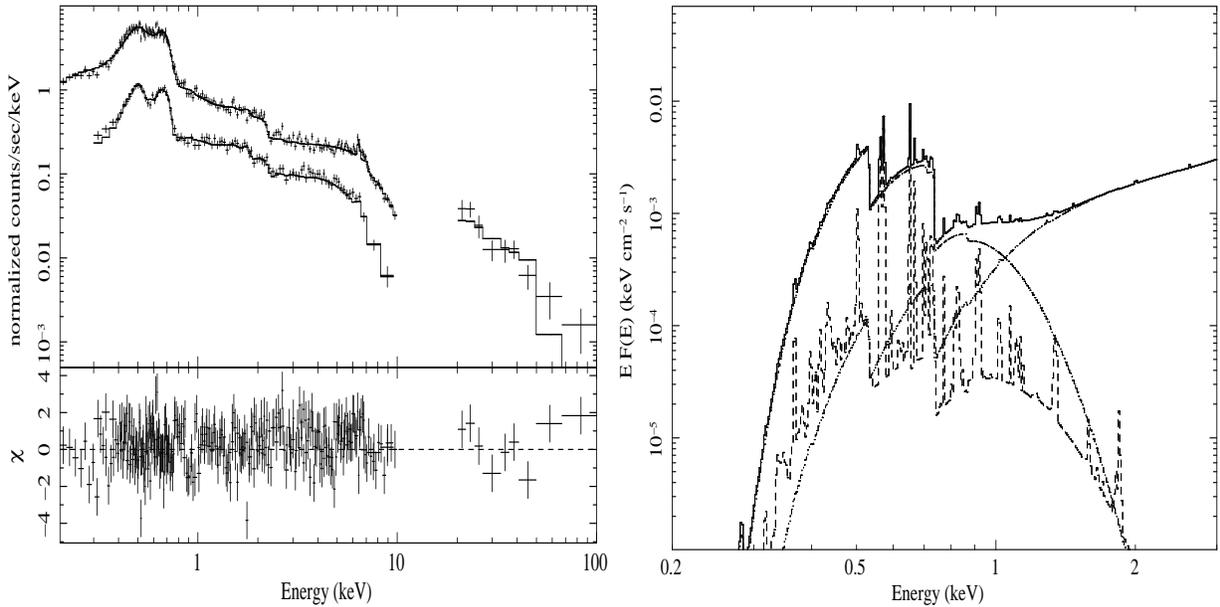

   \centering
\includegraphics[height=8.cm,width=8.cm,angle=-90]{8368_f5a.ps}
\includegraphics[height=8.cm,width=8.cm,angle=-90]{8368_f5b.ps}
\caption{{\it Left:} The EPIC PN (top) and combined MOS (bottom) and IBIS 
spectra fitted simultaneously with the multicomponent model B (see 
text).The lower panel reports the residuals between predicted and observed 
 spectra. {\it Right:} The unfolded model  
including the contribution of various components is plotted up to 
2\,keV, showing the complexities in the soft portion of the 
spectrum.}\label{fig5}
    \end{figure*}

\section{The X-ray spectrum of RX\,J1730}

Identification of  spectral components was performed on the {\em 
XMM-Newton} EPIC PN and the combined MOS averaged spectra  using the  
{\sc XSPEC}   package.  We extracted the spectra between 0.3 and 
10\,keV, outside which the calibration accuracy of the MOS cameras is 
low. Also, because of the non simultaneity of the {\em INTEGRAL} 
observations, the  
{\em INTEGRAL}/ISGRI spectrum was included in the spectral 
analysis at a later stage (see 
Fig.~\ref{fig5} and Table~\ref{spectra}). 

The X-ray spectrum below 1\,keV is 
characterised by a broad feature between 0.4 and 
0.7\,keV followed by a cut-off, whilst  at high 
energies, it shows the well known  iron complex 
including the fluorescent 6.4\,keV line and a hard tail. 
The soft portion of the spectrum is very atypical for IPs (e.g. 
Haberl \& Motch \cite{Haberl95}; Mukai et al. \cite{mukai95}, 
Ramsay et al. \cite{ramsay98},  Haberl et al.\cite{Haberl02}, 
Evans \& Hellier \cite{Evans04},  de Martino et al. \cite{demartino04}, 
\cite{demartino06a},  \cite{demartino06b}, 
Evans et al. \cite{Evans06}).

The EPIC spectra require two optically thin components
as well as a blackbody emission absorbed by multiple components. 
Acceptable fits are achieved including two {\sc mekal} emissions, a 
blackbody and three absorption components. The last consist of 
a total ({\sc wabs}),  a partial ({\sc pcfabs}) absorbers, and an 
absorption {\sc edge}, 
defined as $e^{-\tau(E/E_c)^3}$ for $E\ge E_{c}$, 
where $\tau$ the optical depth at threshold energy $E_{c}$ and 1 for $E\le 
E_{c}$.  With model\,A we then define a composite spectrum: 
{\sc wabs*pcfabs*edge(mekal+bbody+mekal+gaussian)}, where the 
gaussian line accounts for the iron fluorescent line fixed at 6.4\,keV. 
This model gives $\chi^2_{\nu}$=0.995 and in 
 Table~\ref{spectra}, A,  we report the spectral parameters. 
The energy at which this edge is found (0.74\,keV) implies  
{\sc OVII} K-shell absorption with an optical depth at this energy $\tau 
\sim 1.8$. No {\sc OVIII} edge, expected at 0.87\,keV, is observed. 
This absorption feature is indicative of the presence of a warm absorber 
along the line of sight. However, the description with   an  
ionised absorber {\sc absori} (Done et  al. \cite{Done92}) does not 
provide an equally acceptable fit ($\chi^2_{\nu}$=1.09), leaving the gas 
parameters unconstrained. 
The total absorber has an column density  about twice that of the Galactic 
absorption in the direction of 
the source ($\rm N_H = 1.7\times 10^{21}\,cm^{-2}$), indicating an 
intrinsic  contribution of the source. The second intervening column has a 
density  $\rm N_H = 1.4\times 10^{23}\,cm^{-2}$ and covers  $\sim 56\%$ of the 
X-ray source. Multiple absorption components are indeed observed in many X-ray 
spectra of IPs (Mukai et al. \cite{mukai94}). The optically thick 
component is rather hot ($\rm kT_{BB}$=80\,eV) when compared with those 
observed in the polars and the soft X-ray IPs, but it is 
 similar to that found in some hard IPs (Haberl et al. \cite{Haberl02}, 
de Martino et al. \cite{demartino04}, \cite{demartino06a}).  The inclusion 
of the blackbody under only the partial absorber provides a worse 
fit ($\chi^2_{\nu}$=1.04). 
The two optically  thin components are found at 
$\rm kT_1$=0.17\,keV and $\rm kT_2 \geq$ 60\,keV with sub-solar metal 
abundance $\rm A_Z$=0.33. 
A multi-temperature plasma 
({\sc cemekl}) is not required by the data, providing a $\chi^2_{\nu}>$2 
even when including the {\em INTEGRAL}/ISGRI data. This indicates that 
the 
power law temperature dependence of the emission measure is a poor 
description 
of the post-shock region. A similar result was also found  in other 
systems (e.g. Evans \& Hellier \cite{Evans04}, de Martino et al. 
\cite{demartino06b}).  
However, the iron complex in the 6-7\,keV region was inspected 
against  the presence of intermediate temperatures by adopting a hot {\sc 
mekal} at $\rm kT_2$= 60\,keV with $\rm A_Z=0.33$  
plus a gaussian centred  on the 6.4\,keV line and the absorption 
parameters fixed at the values of model\,A).  The 
model predicts the hydrogenic 6.97\,keV iron line 
but, as expected, not the He-like iron line at 6.7\,keV. Inclusion of  
a gaussian line at 6.7\,keV  gives E.W.=44$\pm$ 8\,eV. This emission 
line can be accounted  for by a 10\,keV plasma, suggesting that there 
are regions of the optically thin post-shock plasma  at lower 
temperatures.

The inclusion of the  {\em INTEGRAL}/ISGRI spectrum lowers the hot 
temperature {\sc mekal} to 60\,keV 
(see Table~\ref{spectra}, B). The uncertainty of the instrument 
flux calibration was taken into account by introducing a multiplicative 
constant into the spectral model and normalizing the fluxes to the PN. 
This constant was 0.92. Again the metal
abundance results to be sub-solar $\rm A_Z$=0.40$^{+0.06}_{-0.08}$. 
We remark that it is relative to Anders \& Grevesse 
(\cite{andersgrevesse}) solar abundances that are  higher than 
those adopted for the interstellar medium (e.g. Wilms et al. 
\cite{wilms}, but also see Grevesse et al. \cite{grevesse}). The 
differences between the two are still within errors.
The flux in the 0.2--10\,keV range is $\rm 
1.68\pm0.02\,\times10^{-11}\,erg\,cm^{-2}\,s^{-1}$.
This fit is reported in the left panel of Fig.~\ref{fig5} and in the right
panel we show the unfolded model 
together with the different spectral components.

The high temperature found for the hot optically thin 
component and  the relatively large width (E.W.=110\,eV) of the 
fluorescent $\rm  K_{\alpha}$ iron line at 6.4\,keV could suggest 
illumination  from cold material due to Compton  reflection (Matt et al. 
\cite{matt91}; Done  et al. \cite{Done92}). This component has been  
detected in many magnetic  CVs (Matt et al. \cite{matt}, de Martino et al.  
\cite{demartino01}; Ramsay \& Cropper \cite{ramsay04}) and 
alleviates the problem of high plasma temperatures. 
However, including this component in the fit, 
 (Magdziarz \& Zdziarski \cite{Magdziarz95}) no improvement is found 
($\chi^{2}_{\nu}$=0.98).
We are then left with an uncomfortably high plasma temperature that might 
be regarded as  representative of the immediate post-shock regions of the 
accretion flow.

   \begin{table*}[t]
      \caption{ A: Spectral parameters as derived from  
Ãfitting simultaneously the EPIC PN and MOS phase--averaged spectra
for the  best fit model discussed in the text. Quoted 
errors  refer to 
90$\%$confidence level for the parameter of  interest. B refers
to the inclusion of IBIS/ISGRI spectrum.}
         \label{spectra}
     \centering
\begin{tabular}{ l c c c c c c c c c c c }
            \hline \hline
            \noalign{\smallskip}
            \noalign{\smallskip}

 & $\rm N_H^{a}$  & $\rm N_H^{b}$   & $\rm Cov_{F}^{c}$ & $E_c$ 
& $\tau_{max}$ 
& $\rm kT_{BB}$ & $\rm A_Z^{d}$ & $\rm kT_{1}$ & $\rm kT_{2}$ &
 E.W.$^{e}$ & $\chi^2$/d.o.f.  \cr
            \noalign{\smallskip}
 & & & & (keV) & & (eV) & 
 &  (keV) & (keV) & (eV) & \cr 
            \hline
            \noalign{\smallskip}
{\bf A} & 3.57$^{+0.24}_{-0.25}$ & 1.40$^{+0.32}_{-0.17}$ & 
0.56$^{+0.03}_{-0.02}$ & 0.740$\pm0.005$ & 1.82$^{+0.17}_{-0.16}$ &
 79.5$^{+3.4}_{-3.3}$ & 0.33$^{+0.37}_{-0.19}$ &
 0.174$^{+0.006}_{-0.015}$ & $>59.5$ & 114$\pm$29 & 924/928 \cr
            \noalign{\smallskip}
{\bf B} & 3.11$^{+0.03}_{-0.06}$ & 2.03$^{+0.19}_{-0.30}$ &
0.56$\pm$0.02 & 0.736$^{+0.002}_{-0.006}$ & 1.65$^{+0.07}_{-0.06}$ &  
91.4$^{+0.9}_{-0.8}$ & 0.40$^{+0.067}_{-0.08}$ & 
0.168$^{+0.008}_{-0.003}$ &  60$\pm$6 & 107$\pm$30 & 898/937 \cr
            \noalign{\smallskip}
            \noalign{\smallskip}
            \hline
            \hline
\end{tabular}
~\par
\normalsize
\begin{flushleft}
$^a$: Column density of total absorber in units of 
$10^{21}$~cm$^{-2}$.\par
$^b$: Column density of the partial absorber in 
units of $10^{23}$~cm$^{-2}$.\par
$^c$: Covering fraction of partial absorber.\par
$^d$: Metal abundance in units of the cosmic value (Anders \& Grevesse
\cite{andersgrevesse}) linked for the two thin plasma emissions.\par
$^e$: Equivalent width of gaussian centred at 6.4\,keV.\par  
\end{flushleft}
\end{table*}

\begin{figure}[h!]
   \centering
\includegraphics[height=8.cm,width=8.cm,angle=0]{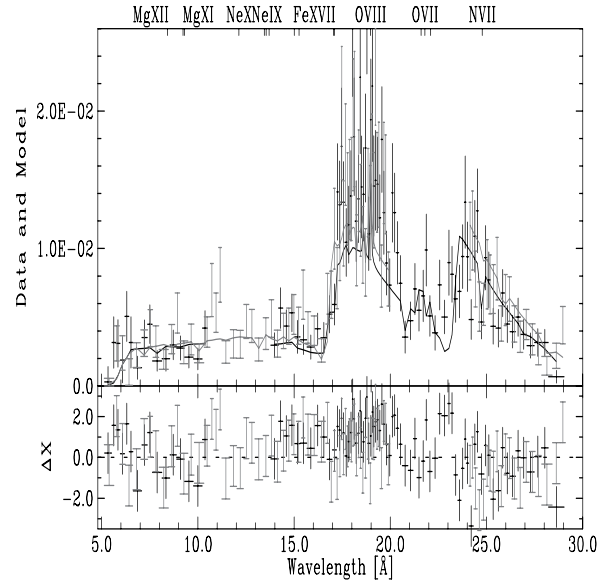}
\caption{The RGS1 (black) and RGS2 (gray)  spectra fitted with model
parameters derived from the combined 
EPIC and ISGRI spectral analysis, but with lower abundance 
($A_Z$=0.21) as discussed in the text. The positions of lines expected to be 
strong are also reported. The residuals  between model and observed 
spectrum are shown in the lower panel.}\label{fig6}
    \end{figure}

 The observed complexity in the 0.6\,keV region and hence the cool 
{\sc mekal} can also be  reproduced  by  a gaussian centred  at 0.568\,keV 
(21.81\,$\AA$)  implying that  {\sc O\,VII} He-like triplet line is 
present. 
We then analysed the RGS spectra applying  the same  model used for 
the EPIC data to  
both RGS1 and RGS2 spectra, but this  predicts too much flux in 
the {\sc 
O\,VII} line. Because of the low data quality, we left free to vary only 
the total absorber and  {\sc mekal} abundance, while keeping all other 
parameters fixed at the values  reported in 
Table~\ref{spectra}, B. The major 
difference was 
found in the abundance, which is lowered to  $A_Z$=0.21. The RGS spectra 
are shown in  Fig.~\ref{fig6} together with the expected positions of 
main 
emission  lines. Worth noticing is the strong {\sc O\,VII} edge at 
$\sim$17\,$\AA$, which is the  main feature of the RGS spectrum and the 
lack of strong 
lines. The {\sc OVIII} (at 19\,$\AA$) line might be present though 
reliable parameters  are difficult to derive due to its weakness
and broadness $\sim$30\,eV. 
 We also included the {\sc absori} component in 
place of the  edge and find that this component is able to reproduce  
the observed spectrum  ("bump") in the 15-20\,$\AA$ region, though also in 
this case we are unable to derive constraints on ionization parameters.  
In summary 
the RGS spectra confirm the presence of a warm absorber in RX\,J1730.

\section{Discussion}

Our {\em XMM-Newton} observation of RX\,J1730 complemented with high 
energy {\em INTEGRAL} data and with optical {\em WHT} photometry
revealed new properties of this intriguing CV. The first detection of 
strong X-ray pulses at the previously identified optical 128\,s period 
confirms that this is one of the most rapidly rotating WDs among IPs
and that its X-ray emission is one of the most complex ever observed.

\subsection{The emission properties}

We detected atypical spectral properties in RX\,J1730. Our combined 
{\em XMM-Newton} and {\em INTEGRAL} analysis has shown that it is a very 
hard source with complex low energy properties. The high energy 
portion of the spectrum is dominated by the hottest post-shock regions of 
up to temperatures of $\sim$50-60\,keV. These values, if  taken as 
representative of the shock itself, would imply $\rm 
M_{WD}=0.89-1.02\,M_{\odot}$ and hence a 
rather  massive primary.   The low energy part of the 
X-ray  spectrum reveals the presence of  absorbed soft components: a
blackbody of similar temperature (90\,eV) as found in 
other hard X-ray IPs observed with {\em XMM-Newton} and {\em 
BeppoSAX} (e.g. Haberl et al. \cite{Haberl02}; de Martino 
et al. \cite{demartino04}, de Martino et al. \cite{demartino06a}), but much 
hotter than that detected in the 
"classical" soft IP systems (30-50\,eV) (Haberl \& Motch \cite{Haberl95}, 
de Martino  et al. \cite{demartino04},\cite{demartino06b}), and a
relatively cool optically thin component at 0.17\,keV. The last 
was also identified  in the {\em XMM-Newton} spectra of the soft systems 
V405\,Aur (Evans \& Hellier \cite{Evans04}), PQ\,Gem (Evans et al. 
\cite{Evans06}) and UU\,Col (de Martino et al. 
\cite{demartino06b}).  This component is further confirmed by the
detection of the {\sc OVII} He-like emission line in the RGS spectrum.
The absorption in RX\,J1730 is very complex, consisting of two high density 
intervening columns with $\rm N_H \sim 3\times 10^{21}\,cm^{-2}$ and $\rm N_H \sim 2\times 
10^{23}\,cm^{-2}$, the last covering $\sim 56\%$ the X-ray source,
as well as of a ionised absorber detected from the K-shell absorption edge 
of 
{\sc O\,VII}. The last has a relatively high optical depth ($\tau \sim 
1.7$). 
 Using the photoionization cross section of {\sc OVII}
(Verner et al. \cite{verner}), we derive a column density
for {\sc O\,VII} of $\rm \sim  7\times 10^{18}\,cm^{-2}$. Assuming that this 
represents all the oxygen 
atoms, the equivalent total hydrogen column, for a solar abundance of oxygen $8.51\times 10^{-4}$ 
(Anders \& Grevesse \cite{andersgrevesse}), is $\rm 8\times 10^{21}\,cm^{-2}$. This value
is  similar to that obtained using the warm absorber model.  
 The material has a relatively high ionization level and might be 
also partially responsible for the red-shifted {\sc 
O\,VII} line due to photoionization. While absorption edges 
are often detected in AGN nuclei and QSOs  (Piconcelli et al. 
\cite{piconcelli04},\cite{piconcelli05}), 
the first unambiguous detection 
of a warm absorber in an IP was found in the {\em Chandra} spectrum of   
V1223\,Sgr (Mukai et al. \cite{mukai01}), where a 
similar {\sc O\,VII} edge was detected.  A second 
detection of an absorption edge, but from ionised iron, was found in the 
IP  V709\,Cas (de Martino et al. \cite{demartino01}) from {\em Rossi XTE} 
observations. As proposed by Mukai et al. 
(\cite{mukai01}), photoionization in the pre-shock flow might occur when 
the shock  height is small compared to the diameter of the flow.  
It could be also occurring in RX\,J1730, as this system could be a weak 
magnetic field accretor (see below).  Due to the lack of adequate
temporal coverage it is not possible to further explore this hypothesis 
via phase--resolved spectroscopy,  since the warm absorber feature should 
be deepest when viewing the  post-shock region through the pre-shock 
flow.

The increasing evidence that IPs also possess a non-negligible, but 
heavily absorbed, soft X-ray blackbody component is a challenging new 
result that {\em BeppoSAX} and {\em XMM-Newton} data  
recently brought into  light. Whether IPs show this component because 
of geometric factors has been recently discussed by Evans \& Hellier 
(\cite{Evans07}). 
The bolometric blackbody flux $\rm  
F_{BB}=1.31\times10^{-10}\,erg\,cm^{-2}\,s^{-1}$, when 
compared with that of the optically thin components 
$\rm F_{thin}=1.5\times10^{-10}\,erg\,cm^{-2}\,s^{-1}$, yields a 
 ratio of blackbody to plasma emission of $87\%$. Comparing this ratio 
with those found in other IPs 
 (Haberl et al. \cite{Haberl02}; de Martino et al. 
\cite{demartino04}, \cite{demartino06a}, \cite{demartino06b}), there appears 
to be a tendency for higher ratios for hotter blackbodies, 
reflecting the strong dependence on the temperature.
 These high temperatures could arise from small "cores" of the heated 
magnetic polar areas on the WD surface.  
 In RX\,J1730 the blackbody bolometric luminosity 
$\rm L_{BB}=1.6\times10^{34}\,d_{1kpc}^2\,erg\,s^{-1}$   is very 
high. This gives an emitting area  $\rm a_{BB}=2.3\times 
10^{14}\,d_{1kpc}^2\,cm^2$ and a fractional WD area $\rm f \sim 3\times 
10^{-5}\,d_{1kpc}^2$.  Though this parameter 
is subject to model uncertainties, it is by three orders of magnitude 
lower   than those derived in the soft X-ray IPs showing cooler 
blackbodies such 
as PQ\,Gem (James et al. \cite{jamesetal}). However,  
it is larger by one order of magnitude than that in  NY\,Lup  
that shows a similar hot blackbody component 
(Haberl et al. \cite{Haberl02}).

The emission measure of the optically thin plasma,  $\rm E.M.\sim 
24\times 
10^{55}\,d_{1kpc}^2\,cm^{-3}$, also  turns out to be large if the system 
is at 
1\,kpc. This  might favour high densities as also suggested by the 
large column densities and high mass accretion rate (see below).

 As far the distance is concerned,  G\"ansicke 
et al. (\cite{Gaensicke05})  estimate a donor star of type G0V-G6V contributing 
 15$\%$ of the  total  flux at 6800\,$\AA$ and a 
reddening $\rm  E_{B-V}$=0.45.  This reddening is consistent with the 
hydrogen column 
density derived for the total absorber from the X-ray spectral analysis.
They also derive an amplitude of 115\,km\,s$^{-1}$ from He\,II emission line 
radial velocity curve. Assuming that this represents the motion of the 
primary star in a 15.4\,hr orbit, we infer a mass function $\rm 
f(M_{sec}$)=0.101\,M$_{\odot}$.
Using the WD mass estimate  
0.89-1.02\,$\rm M_{\odot}$, the secondary mass would 
then be in the range  $\rm M_{sec}$=0.62-0.92\,$\rm M_{\odot}$ for 
inclination angles $i \sim  50-90^{\circ}$. Lower 
values of the 
inclination angle ($i<50^{\circ}$) would  imply a donor star more 
massive than the primary.  Further constraints on the 
system geometry are derived below.
The secondary radius,  assuming a Roche-lobe filling star, would be in the 
range $\sim $1.2-1.4$\rm R_{\odot}$ for a 15.4\,hr orbital period. Because 
of the dominant contribution of accretion to the optical  flux, 
 we use the infrared  K band measure from 
2MASS survey reported in G\"ansicke et al. (\cite{Gaensicke05}) to derive 
a   lower limit to the distance of RX\,J1730. This because  
the  secondary  star is expected to dominate the IR flux, although it 
cannot be  excluded that cyclotron emission, outer disc or circumbinary 
cool material (Taam \& Spruit \cite{taamspruit}) could 
 contribute to the K band flux.  Hence, 
 adopting a G0-G6 type donor star with $\rm R_{sec}$ = 1.0-1.4\,$\rm 
R_{\odot}$, the reddening reported above, and using the K band surface 
brightness, we derive $\rm d \geq$ 1.0-1.6\,kpc. 
Though it is the minimum distance, the assumed value of 8\,kpc by 
Masetti et al. (\cite{masetti04}) would imply a bolometric luminosity 
much larger (by two orders of magnitude) than typical values of CVs in 
quiescence. 

The total luminosity, including the soft and hard components 
results to be 
$\rm L_{BOL}=3.36\times10^{34}\,d_{1kpc}^2\,erg\,s^{-1}$.
This 
translates into a mass accretion rate of  $\rm \sim 
2.2-2.5\times10^{-9}\,d_{1kpc}^2\,M_{\odot}\,yr^{-1}$ for $\rm 
M_{WD}=0.89-1.02\,M_{\odot}$.  While this is a  high value if 
the source is indeed at a distance of 1\,kpc, it is not unexpected for its 
long orbital period. To this regard we note that the average secular 
mass accretion rate for systems above the orbital period gap is $\rm \sim 
10^{-9}\,M_{\odot}\,yr^{-1}$ (McDermott  \& Taam \cite{mcdermott}), as 
the result of magnetic braking of the secondary star that acts as angular 
momentum loss mechanism in these systems.  As a comparison,  
the long period  IP  NY\,Lup ($\rm P_{orb}$=9.87\,hr) was found at 
a mass accretion rate of $\rm 
\sim  8\times 10^{-10}\,M_{\odot}\,yr^{-1}$ (de Martino et al. 
\cite{demartino06c}). At a similar orbital period, the peculiar AE\,Aqr 
($\rm P_{orb}$=9.88\,hr), is not in an 
accretion regime but in a propeller state with a secondary transferring 
mass at a  rate of  $\rm \sim 6.4 \times 10^{-9}\,M_{\odot}\,yr^{-1}$ 
(see also Venter \& Meintjes \cite{venter}). Since, RX\,J1730 shows
all signatures of an accreting magnetic WD, the estimated rate is 
not too far from what expected from magnetic braking. 

Furthermore, accretion from the stellar wind of the G-type donor 
star is not expected to be the main source  of the observed X-ray 
luminosity. This because adopting  a power law 
dependence of the mass loss rate on the surface magnetic field strength 
(Collier Cameron \cite{colliercameron}) and a secondary star surface 
magnetic field of 115\,G, derived from the donor rotational (orbital) 
period, a wind  mass loss rate of $\rm \sim 5\times 
10^{-11}\,M_{\odot}\,yr^{-1}$ is estimated.
Indication of a high accretion rate also comes from the large contribution 
of  the accretion flow in the red band (G\"ansicke 
et al. \cite{Gaensicke05}) where, instead, long-period systems 
are less affected. It might also favour an extended accretion 
disc consistent with the wide orbit of the system. The condition for 
disc formation and truncation at the magnetospheric boundary is 
$\rm R_{mag}\sim R_{co}$, where $\rm 
R_{co}=(G\,M_{WD}\,P_{\omega}^2/4\,\pi^2)^{1/3} \sim 3.7\times
10^9\,cm$ is the corotation radius at which
the magnetic field rotates with the same Keplerian frequency of the WD and 
where $\rm R_{mag}= 
5.5 \times 10^{8}\, 
(M_{WD}/M_{\odot})^{1/7}\,R_{9}^{-2/7}\,L_{33}^{-2/7}\,\mu_{30}^{4/7}\,cm$  
is the magnetospheric radius, where  $\rm R_{9}$ is the WD radius in 
units of $10^{9}$\,cm, $\rm
L_{33}$ is the luminosity in units of $\rm 10^{33}\,erg\,s^{-1}$, and
$\mu_{30}$ is the WD magnetic moment in units of $\rm 10^{30}$\,G\,cm$^3$.
From the derived accretion  luminosity and WD mass, we estimate a 
magnetic moment  $\rm \mu \sim 1.2\times10^{32}\,d_{1kpc}\,G\,cm^{3}$. 
Hence, at a distance of 1\,kpc, RX\,J1730 would host  a weak magnetic 
field WD, consistent with the fast rotation and pulsation 
properties of this source  (see also below). Here we remind that also 
Norton et al. (\cite{norton04}) arrive to this conclusion for the rapid 
rotators in IPs.

\subsection{The white dwarf rotation}

We detected for the first time rapid X-ray pulses at the previously 
identified optical 128\,s period, thus confirming this system as one of 
the most rapidly rotating WD IP. The X-ray pulses are highly 
structured and characterised by a broad and deep principal minimum  and 
a much weaker secondary one. The principal minimum is broadly in phase 
with that observed in the optical modulation  on the same day of the X-ray 
observation. There is a clear
change in amplitudes and shapes moving from high energies (X-rays) to the 
optical (red) bands,  while simultaneous observations in the far-UV 
range do not reveal a detectable modulation.
The increase in amplitude of the X-ray pulsation with decreasing energy
is a typical characteristics of IPs and consistent with 
photoelectric absorption in the accretion flow. The spin pulse changes 
from single-peaked above 5\,keV to a double-peaked  rather structured 
curve below 5\,keV. Phasing and amplitudes suggest that 
accretion  occurs onto two 
opposite polar regions (offset by $180^{\circ}$), with one pole (the 
primary) dominating in the hard X-rays.
 The   other pole (the secondary) is 
essentially soft being $\sim$1.5 times less strong than the  primary 
pole.  The highly  structured  shape of the spin modulation  in the 
softer bands also suggests the presence of absorption effects especially 
at those phases when the secondary pole contributes most (i.e. 
$\rm \phi_{rot}\sim$0.2).

The strong X-ray and optical spin  pulses at the 128\,s period, 
indicates that accretion occurs   predominantly via a disc that  
feeds both poles. Although we observe strong changes in the amplitude of 
the optical pulse from night-to-night, which is not uncommon in IPs, 
the stability of pulses  and  the absence of a beat variability at a 
period $\rm P_{\omega-\Omega}$=128.3\,s in our unpublished 
long multi-epoch photometric optical monitoring supports the accretion 
disc configuration.

 In the accretion curtain  geometry (Rosen et al. \cite{rosen}) the 
material flows from the  accretion disc towards the  magnetic poles of the 
WD in an arc-shaped  curtain, where the optical depth  is larger along the 
field lines. This  gives rise to  X-ray and optical 
maxima that are in phase when the curtain points away from the observer, 
while a minimum is observed when the polar region point towards the observer. 
The double-peaked shape and the phasing of X-ray and optical light 
curves suggest that this is the case, provided that the geometry allows 
the two opposite poles to come into view to the observer  with 
different contributions. This in turn implies that the binary
inclination, {\em i}, and magnetic colatitude, {\em m}, assume specific 
ranges of values. 

Constraints on the binary inclination 
come from the lack of  eclipses, cos\,$(i) \rm > (R_{WD} + 
R_{sec})/a$ 
where  a is the binary separation ($\sim 3.6\,R_{\odot}$ from the mass 
estimates and orbital period), giving $ i \leq 70^{\circ}$. The presence 
of  ellipsoidal  variations in 
the optical light (G\"ansicke et al. \cite{Gaensicke05}) is also 
consistent with this angle. 
As discussed above, $i > 50^{\circ}$, thus limiting $ 50^{\circ} < i < 
70^{\circ}$. 
Furthermore, the 
visibility of the two poles implies $ |i - m| < 90^{\circ} - \beta$, where 
$\beta$ is  the subtending angle of the polar cap, which is related to 
the  fractional accretion area by cos\,$\beta = 1 -2\,f$ (King \& Shaviv 
\cite{kingandshaviv}). Since $f<<1$, $\beta \sim 2\,f^{0.5}$ 
and it can be neglected at first approximation.
Because the optical pulse shows similar maxima, the two poles should 
have similar visibility,  implying that the magnetic colatitude is 
moderately large (see also Evans \& Hellier \cite{Evans07}). Absorption 
effects  are observed in this system, suggesting $m \leq i$. 

\begin{figure}[h!]
   \centering
\includegraphics[height=5.cm,width=8.cm,angle=0]{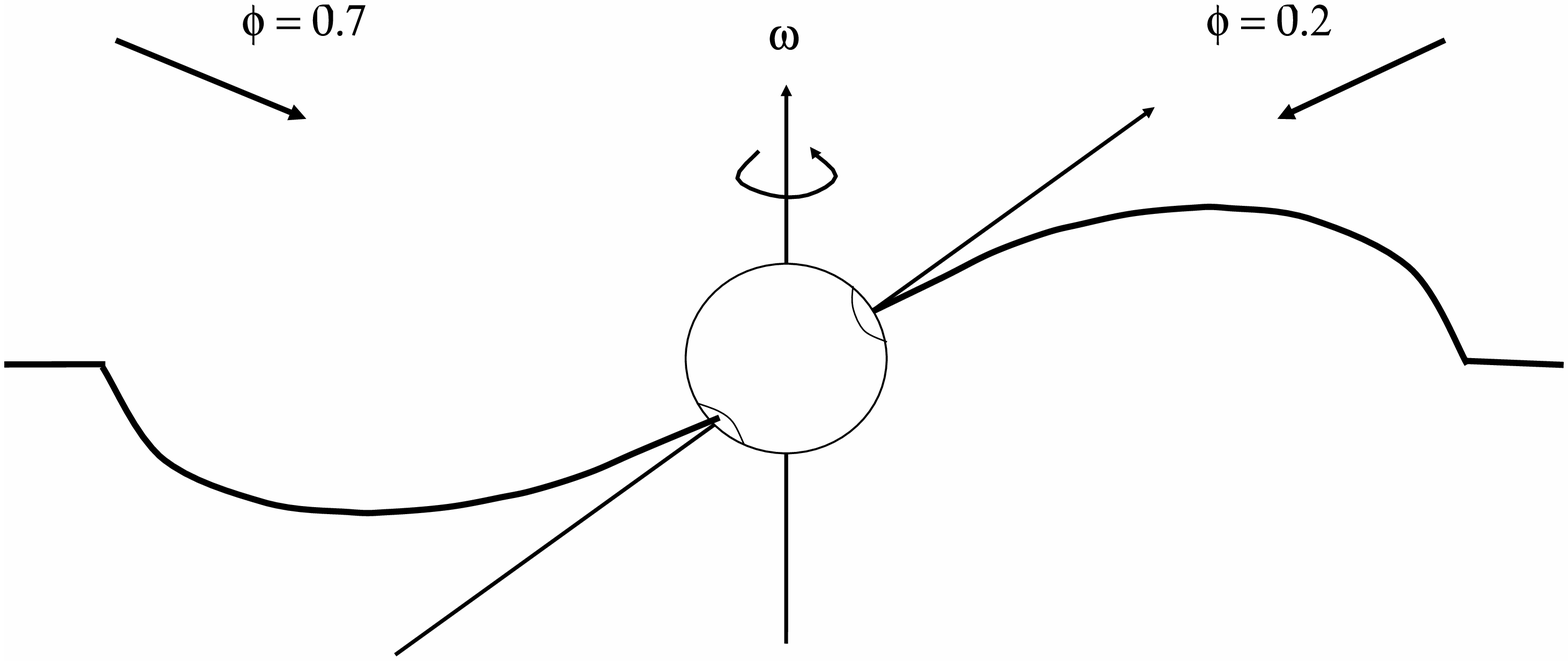}
\caption{A schematic picture of the accretion geometry in RX\,J1730. A 
binary inclination $i=65^{\circ}$ and a magnetic colatitude $m=45^{\circ}$ 
are adopted. The phases at which the two maxima of the spin pulses occur 
are also denoted.}\label{fig7}
    \end{figure}

A schematic picture of the 
accretion flow is shown  in Fig.~\ref{fig7} 
 for an inclination angle $i =65^{\circ}$ and 
magnetic colatitude $m= 45^{\circ}$.  In this configuration the lower pole
is better represented and contributes most at rotational phase $\sim$0.7
when the visibility of the lower curtain is maximum. 
This pole (primary) is responsible for the maximum spin pulse at 
this phase at 
all energies (from hard X-rays to the optical). On the other hand, the 
upper pole is less represented (secondary pole) and, at rotational phase 
$\sim$0.2, it is  affected by absorption as indeed detected. At this 
phase the  heated polar region of the WD atmosphere is at its maximum 
visibility, hence contributing most in the soft X-rays as observed. 
The magnetospheric radius extends up to $\rm \sim 
6\,R_{WD}$ suggesting that field lines reconnect far from the WD. If the 
magnetic colatitude is moderately large, the curtain tends to flatten 
in the orbital plane, allowing the outer regions of the upper curtain 
to be also visible and hence to contribute to the spin pulse at rotational 
phase 0.2. The inner regions of the upper curtain would instead  
contribute little at this phase. The asymmetry in the optical 
pulse at spin phase $\sim$0.35,  best seen at red wavelengths, could be 
due to the twisting of field lines due to the  
fast WD rotation, which allows the 
upper curtain to also contribute at other phases. 

The spin pulse would then be due to the combination of different 
contributions with likely different spectral 
properties and difficult to isolate. This is known to be the case in 
those IPs that were 
studied  in a wide wavelength range (e.g.
Welsh \& Martell \cite{Welsh}, de Martino et  al. \cite{demartino99},
Eisenbart et al. \cite{Eisenbart}). The pulses in these systems were found 
to be  multi-component, ranging from the hot WD pole or inner regions of 
the  curtain to the cooler outer regions of the curtain itself. 

In RX\,J1730, the pulse appears to peak in the g' band. To obtain 
information on the overall spectral shape of the pulse, we used the 
UVW1 flux, the approximate average magnitudes in u', g', and r' bands
(using colour relations by Fukugita et al. \cite{Fukugita}) and the 
derived spin amplitudes.  Using  Fukugita et al. 
Sloan zero 
points and $\rm E_{B-V}$=0.45, we obtained dereddened modulated fluxes of
$\rm F_{UVW1}\sim 1.1\times 10^{-15}\,erg\,cm^{-2}\,s^{-1}$, $\rm 
F_{u'}\sim 1.7\times 10^{-15}\,erg\,cm^{-2}\,s^{-1}$, $\rm F_{g'}\sim 
1.5\times 10^{-15}\,erg\,cm^{-2}\,s^{-1}$ and $\rm F_{r'}\sim 3.7\times 
10^{-16}\,erg\,cm^{-2}\,s^{-1}$. These indicate a blackbody 
colour temperature of 6800--7800\,K,  except for the r' band
whose flux is too low for any blackbody shape. Since we do not have 
standard star optical photometry, the  above fluxes should be regarded as 
rough estimates. 
Furthermore, the extrapolation of the unabsorbed soft 
X-ray  blackbody component to the UVW1 band predicts a flux 
that is 4 orders of magnitude lower than the dereddened observed one. 
Hence, the spin pulse appears to be  dominated by relatively cool 
region(s) of the curtain. Although we cannot exclude the 
presence of the irradiated WD poles in the UV, these would be 
masked by the Ãlarge   curtain contribution. 
Here we remark that also the  soft IP PQ\,Gem shows an extremely weak UV 
spin modulation (Evans et al. 
\cite{Evans06}), while others like UU\,Col has a strong UV spin pulsation 
dominated by the heated WD polar regions  (de Martino et al. 
\cite{demartino06b}). 

Evans \& Hellier (\cite{Evans07}) discussed the different 
behaviour of IPs in terms of dipole inclinations: those soft X-ray 
systems, like  UU\,Col and V405\,Aur,  showing double humped light curves 
are also found to be less affected by absorption because of 
highly inclined magnetic axis. RX\,J1730 could be 
an intermediate case where the magnetic dipole is moderately  
inclined.
Furthermore, by analogy with these soft X-ray IPs
that show two poles,  and the polars, in which the 
secondary accreting pole is softer than the main pole, we might 
argue that RX\,J1730 has a similar property. Interestingly, the 
secondary  pole in the polars is found to possess a stronger field than 
the primary pole (Beuermann \cite{Beuermann99}).

\subsection{The accretor in RX\,J1730}

 Fast rotational periods  of the accreting WD and double-peaked X-ray 
pulses were claimed to be a common feature of  relatively  weak magnetic 
field accretors (Norton et al.  \cite{norton99};  
\cite{norton04}). The very rapid WD rotation, 
the complex X-ray  pulse and the low magnetic moment of RX\,J1730 
(even at a distance larger than 1\,kpc), 
might then add this system to the  group of low magnetic field systems.
Either  wide or tall accretion regions could produce a double-peaked 
pulse profile (Norton et al. \cite{norton99}). This is 
difficult to establish in general and in particular in RX\,J1730, 
because no spin phase--resolved  X-ray spectral analysis can be performed. 
However, the presence of a warm absorber might favour wide accretion spots 
as  proposed by Mukai et al (\cite{mukai01}). The appearance of 
double-humped 
spin light curves and soft X-ray component can be also produced by highly 
inclined dipole fields (Evans \& Hellier \cite{Evans07}). To explain the 
observed properties of RX\,J1730, 
the WD magnetic field axis is 
likely to be moderately inclined and, if indeed the WD is weakly 
magnetised, this  system will not be prone to synchronism (Norton 
et al. \cite{norton04}). 

Furthermore, the WD is rapidly rotating but less rapidly than  AE\,Aqr and 
it is neither spinning-down   as instead AE\,Aqr does, nor  spinning-up.
It could be possible that  RX\,J1730 is a young 
system born with a rapidly rotating WD. Very rapid rotations could be 
the result of a thermal timescale mass transfer phase 
as proposed by Schenker et al. (\cite{schenker02}) for AE\,Aqr itself 
and other CVs that show peculiar spectral  features of CNO processing. 
These systems could be descendant from 
supersoft X-ray binaries that had formerly massive secondary stars
and run mass transfer on a thermal timescale that allowed 
nuclear stable burning onto the WD. At the end of the brief supersoft 
phase the systems evolve off to longer orbital period or switch to stable 
mass transfer. They also predict that about 
one-third of CVs have undergone  such evolution.  Signatures of 
non-standard evolution are spectral  features due to CNO processing due 
to the chemical evolution of the donor stellar interior. These 
are usually observed in the UV range  through anomalous line intensities 
of Carbon, Nitrogen and Oxygen (see also G\"ansicke et al. 
\cite{gaensicke03}). Although we lack of UV spectroscopy to corroborate 
this possibility, this appears to be an interesting scenario that deserves 
further investigation. Also, a precise 
characterisation of the 
secondary  star with adequate optical and nIR observations is needed 
to further progress in the understanding of this 
system.

RX\,J1730 was also recently claimed to be detected at radio 
wavelengths with a very likely non-thermal emission and was 
proposed to be a LMXB (Pandey et  al. \cite{pandey06}). 
A neutron star (NS) nature and hence a LMXB was also proposed by 
Masetti  et al. (\cite{masetti04}) from the high luminosity that would 
result from locating this source in the galactic bulge.
However we remark that the radio 
VLA survey position even including the NVSS catalogue errorbars  (Condon 
et al. \cite{condon98}), while it is  within the 
{\em INTEGRAL} error circle, does not agree with the optical position 
(G\"ansicke et al. \cite{Gaensicke05})  and it is outside the ROSAT error 
circle.  Hence, the claim of the radio detection at 1.4\,GHz should be 
taken with great caution and  further observations are essential to assess 
this issue. 
We also note that very recent radio surveys 
 (Mason \& Gray \cite{mason07}) are  bringing new detection of CVs in this 
range. It is argued that the condition to produce radio emission is that 
secondaries have  magnetic fields of a few hundred to a few thousand Gauss 
(Mason \& Gray \cite{mason07}, Meintjes \& Jurua \cite{meintjes}). It is  
 further suggested that the systems should be disc-less because the few 
CVs identified in the radio regime are polars and pre-CVs.  The only 
IP with a secure radio counterpart  again is   the peculiar system 
AE\,Aqr. Though the fast rotation and  the long orbit of RX\,J1730  
might bring similarities to that system,  AE\,Aqr  is 
a  strongly flaring  source at all wavelengths and it is quite 
atypical in its X-ray emission, being a soft X-ray source with a very 
low accretion rate (see Itoh et al. \cite{itoh} and references therein). 
From the detection of a spin-down of the WD, it is now well established 
that this system is in  a propeller state (Schenker et al. 
\cite{schenker02} and references therein.)
RX\,J1730 is not observed to be a flaring source and its spin period is stable at the quoted 
value.
The analysis of the {\em XMM-Newton} and  {\em 
INTEGRAL} data reveals many similarities with other  IPs. It is a 
very  hard X-ray source, though, in  some respect, peculiar in its soft 
emission. The timing analysis shows  that RX\,J1730 is modulated at the WD spin period up to 
10\,keV, implying that the primary is 
experiencing accretion via a disc. 
 It is therefore difficult to match the observed
characteristics of RX\,J1730 with either a state similar to 
AE\,Aqr or a neutron star accretor.

\section{Conclusions}

We  presented the first  X-ray study of RX\,J1730 based on 
an {\em XMM-Newton} observation, complemented with {\em INTEGRAL}  
 and optical fast photometric data that confirm this CV as a 
magnetic  system of the IP type. 

The rapid 128\,s pulsation is energy dependent with 
remarkable differences from the X-rays, the UV to the optical ranges. We 
identified two emitting poles with different contributions.
A  main emitting pole contributes from soft to the hard X-rays, 
while the secondary pole contributes most in the soft bands. 
 The WD magnetic 
dipole field should be moderately inclined, thus allowing the visibility 
of both poles, with the lower pole better represented. 
The optical spin pulse is mainly due 
to  reprocessing in the accretion flows onto the WD and dominated by the 
outer regions. The lack of 
detectable UV pulsation is an unexpected property, which might indicate 
that reprocessing at both the WD surface and in the inner 
regions of the accretion flow is masked by the contribution of the outer  
flow regions.

X-ray spectral analysis reveals a hard X-ray spectrum as well as 
remarkable soft X-ray complexities.
 These can be identified in a warm  
absorber, which makes this system the second IP where an absorption
 edge from ionised oxygen has 
been revealed, together with multiple optically thin emissions and a 
blackbody  component. The last adds RX\,J1730 to the small group of 
soft X-ray IPs. We however find a large luminosity for this component that 
is 
difficult to explain. The observed X-ray properties are 
consistent with a system hosting a weakly magnetised WD rather than a NS. 
 The long 15.4\,hr orbital period and rapid 128\,s WD rotation 
could suggest that RX\,J1730 is a young IP born with a rapidly rotating 
WD. 
To understand its evolutionary  status, this system deserves further 
observations especially  addressing the donor star parameters.

 \begin{acknowledgements}
DDM acknowledges financial support by  the Italian Space Agency 
(ASI) under contract I/023/05/0 and thanks R. Gonzalez-Riestra at ESAC for 
her help in checking operational performance of the OM during the 
observation.  ULTRACAM is supported by PPARC grant PP/D002370/1. The WHT 
is operated on the island of La Palma by the Isaac Newton Group in the 
Spanish Observatorio del Roque de los Muchachos of the Instituto de 
Astrofísica de Canarias.
\end{acknowledgements}

\end{document}